\pdfoutput=1
\documentclass[12pt,a4paper]{article}

\usepackage{ifthen} 
\newboolean{pdflatex}
\setboolean{pdflatex}{true} 

\newboolean{articletitles}
\setboolean{articletitles}{true} 

\newboolean{uprightparticles}
\setboolean{uprightparticles}{false} 

\def\paperauthors{LHCb collaboration} 
\def\paperasciititle{Search for D*(2007)0 -> mu+ mu- in B- -> pi- mu+ mu- decays} 
\def\papertitle{Search for \decay{\Dstar(2007)^0}{\mumu} in \decay{\Bm}{\pim\mumu} decays} 
\def\paperkeywords{{High Energy Physics}, {LHCb}} 
\def\papercopyright{\the\year\ CERN for the benefit of the LHCb collaboration} 
\def\paperlicence{CC BY 4.0 licence}
\def\paperlicenceurl{https://creativecommons.org/licenses/by/4.0/}

\usepackage[top=1in, bottom=1.25in, left=1in, right=1in]{geometry}

\usepackage{microtype}
\usepackage{lineno}  
\usepackage{xspace} 
\usepackage{caption} 

\usepackage{graphicx}  
\usepackage{color}
\usepackage{colortbl}
\graphicspath{{./figs/}} 

\usepackage{amsmath} 
\usepackage{amssymb}
\usepackage{amsfonts}
\usepackage{upgreek} 

\newcommand*\patchAmsMathEnvironmentForLineno[1]{%
\expandafter\let\csname old#1\expandafter\endcsname\csname #1\endcsname
\expandafter\let\csname oldend#1\expandafter\endcsname\csname
end#1\endcsname
 \renewenvironment{#1}%
   {\linenomath\csname old#1\endcsname}%
   {\csname oldend#1\endcsname\endlinenomath}%
}
\newcommand*\patchBothAmsMathEnvironmentsForLineno[1]{%
  \patchAmsMathEnvironmentForLineno{#1}%
  \patchAmsMathEnvironmentForLineno{#1*}%
}
\AtBeginDocument{%
\patchBothAmsMathEnvironmentsForLineno{equation}%
\patchBothAmsMathEnvironmentsForLineno{align}%
\patchBothAmsMathEnvironmentsForLineno{flalign}%
\patchBothAmsMathEnvironmentsForLineno{alignat}%
\patchBothAmsMathEnvironmentsForLineno{gather}%
\patchBothAmsMathEnvironmentsForLineno{multline}%
\patchBothAmsMathEnvironmentsForLineno{eqnarray}%
}

\usepackage{hyperxmp}

\usepackage[pdftex,
            pdfauthor={\paperauthors},
            pdftitle={\paperasciititle},
            pdfkeywords={\paperkeywords},
            pdfcopyright={Copyright (C) \papercopyright},
            pdflicenseurl={\paperlicenceurl}]{hyperref}

\usepackage[bottom,flushmargin,hang,multiple]{footmisc}

\usepackage[all]{hypcap} 

\usepackage{xspace} 
\usepackage{upgreek}

\def\lhcb   {\mbox{LHCb}\xspace}

\def\MagUp {\mbox{\em Mag\kern -0.05em Up}\xspace}

\ifthenelse{\boolean{uprightparticles}}%
{

 \def\Pmu         {\ensuremath{\upmu}\xspace}                 
 \def\Pnu         {\ensuremath{\upnu}\xspace}                 
                  
 \def\Ppi         {\ensuremath{\uppi}\xspace}                 
                  
 \def\Prho        {\ensuremath{\uprho}\xspace}                 
                  
 \def\Ptau        {\ensuremath{\uptau}\xspace}

 \def\Ppsi        {\ensuremath{\uppsi}\xspace}

 \def\PDelta      {\ensuremath{\Delta}\xspace}                 
 \def\PXi         {\ensuremath{\Xi}\xspace}                 
 \def\PLambda     {\ensuremath{\Lambda}\xspace}                 
 \def\PSigma      {\ensuremath{\Sigma}\xspace}                 
 \def\POmega      {\ensuremath{\Omega}\xspace}                 
 \def\PUpsilon    {\ensuremath{\Upsilon}\xspace}
 \let\oldPi\Pi
 \def\PPi         {\ensuremath{\oldPi}\xspace}

 \def\PB      {\ensuremath{\mathrm{B}}\xspace}                 
                  
 \def\PD      {\ensuremath{\mathrm{D}}\xspace}

 \def\PJ      {\ensuremath{\mathrm{J}}\xspace}                 
 \def\PK      {\ensuremath{\mathrm{K}}\xspace}

 \def\Pb      {\ensuremath{\mathrm{b}}\xspace}                 
 \def\Pc      {\ensuremath{\mathrm{c}}\xspace}                 
                  
 \def\Pe      {\ensuremath{\mathrm{e}}\xspace}

 \def\Pi      {\ensuremath{\mathrm{i}}\xspace}

 \def\Ps      {\ensuremath{\mathrm{s}}\xspace}

 \def\thebaroffset{0.0em}
}
{

 \def\Pmu         {\ensuremath{\mu}\xspace}                 
 \def\Pnu         {\ensuremath{\nu}\xspace}                 
                  
 \def\Ppi         {\ensuremath{\pi}\xspace}                 
                  
 \def\Prho        {\ensuremath{\rho}\xspace}                 
                  
 \def\Ptau        {\ensuremath{\tau}\xspace}

 \def\Ppsi        {\ensuremath{\psi}\xspace}                 
                  
 \mathchardef\PDelta="7101
 \mathchardef\PXi="7104
 \mathchardef\PLambda="7103
 \mathchardef\PSigma="7106
 \mathchardef\POmega="710A
 \mathchardef\PUpsilon="7107
 \mathchardef\PPi="7105
                  
 \def\PB      {\ensuremath{B}\xspace}                 
                  
 \def\PD      {\ensuremath{D}\xspace}

 \def\PJ      {\ensuremath{J}\xspace}                 
 \def\PK      {\ensuremath{K}\xspace}

 \def\Pb      {\ensuremath{b}\xspace}                 
 \def\Pc      {\ensuremath{c}\xspace}                 
                  
 \def\Pe      {\ensuremath{e}\xspace}

 \def\Pi      {\ensuremath{i}\xspace}

 \def\Ps      {\ensuremath{s}\xspace}

 \def\thebaroffset{0.18em}
}
\newcommand{\offsetoverline}[2][\thebaroffset]{\kern #1\overline{\kern -#1 #2}}%

\makeatletter
\ifcase \@ptsize \relax
  \newcommand{\miniscule}{\@setfontsize\miniscule{4}{5}}
\or
  \newcommand{\miniscule}{\@setfontsize\miniscule{5}{6}}
\or
  \newcommand{\miniscule}{\@setfontsize\miniscule{5}{6}}
\fi
\makeatother

\DeclareRobustCommand{\optbar}[1]{\shortstack{{\miniscule (\rule[.5ex]{1.25em}{.18mm})}
  \\ [-.7ex] $#1$}}

\def\epem       {{\ensuremath{\Pe^+\Pe^-}}\xspace}

\def\mup        {{\ensuremath{\Pmu^+}}\xspace}
\def\mun        {{\ensuremath{\Pmu^-}}\xspace} 

\def\mumu       {{\ensuremath{\Pmu^+\Pmu^-}}\xspace}

\def\tautau     {{\ensuremath{\Ptau^+\Ptau^-}}\xspace}

\def\neu        {{\ensuremath{\Pnu}}\xspace}
\def\neub       {{\ensuremath{\overline{\Pnu}}}\xspace}
\def\neum       {{\ensuremath{\neu_\mu}}\xspace}
\def\neumb      {{\ensuremath{\neub_\mu}}\xspace}

\def\squark    {{\ensuremath{\Ps}}\xspace}

\def\cquark    {{\ensuremath{\Pc}}\xspace}

\def\bquark    {{\ensuremath{\Pb}}\xspace}

\def\pion   {{\ensuremath{\Ppi}}\xspace}

\def\pip    {{\ensuremath{\pion^+}}\xspace}
\def\pim    {{\ensuremath{\pion^-}}\xspace}

\def\rhomeson {{\ensuremath{\Prho}}\xspace}
\def\rhoz     {{\ensuremath{\rhomeson^0}}\xspace}

\def\kaon    {{\ensuremath{\PK}}\xspace}
\def\Kbar    {{\ensuremath{\offsetoverline{\PK}}}\xspace}

\def\KorKbar {\kern \thebaroffset\optbar{\kern -\thebaroffset \PK}{}\xspace}

\def\Km      {{\ensuremath{\kaon^-}}\xspace}

\def\Kstarzb {{\ensuremath{\Kbar{}^{*0}}}\xspace}

\def\D       {{\ensuremath{\PD}}\xspace}

\def\DorDbar {\kern \thebaroffset\optbar{\kern -\thebaroffset \PD}\xspace}
\def\Dz      {{\ensuremath{\D^0}}\xspace}

\def\Dp      {{\ensuremath{\D^+}}\xspace}
\def\Dm      {{\ensuremath{\D^-}}\xspace}

\def\DpDm    {\ensuremath{\Dp {\kern -0.16em \Dm}}\xspace}
\def\Dstar   {{\ensuremath{\D^*}}\xspace}

\def\Dstarz  {{\ensuremath{\D^{*0}}}\xspace}

\def\Dstarp  {{\ensuremath{\D^{*+}}}\xspace}

\def\B       {{\ensuremath{\PB}}\xspace}
\def\Bbar    {{\ensuremath{\offsetoverline{\PB}}}\xspace}

\def\BorBbar {\kern \thebaroffset\optbar{\kern -\thebaroffset \PB}\xspace}
\def\Bz      {{\ensuremath{\B^0}}\xspace}
\def\Bzb     {{\ensuremath{\Bbar{}^0}}\xspace}
\def\Bd      {{\ensuremath{\B^0}}\xspace}

\def\BdorBdbar {\kern \thebaroffset\optbar{\kern -\thebaroffset \Bd}\xspace}

\def\Bub     {{\ensuremath{\B^-}}\xspace}

\def\Bm      {{\ensuremath{\Bub}}\xspace}

\def\Bs      {{\ensuremath{\B^0_\squark}}\xspace}
\def\Bsb     {{\ensuremath{\Bbar{}^0_\squark}}\xspace}
\def\BsorBsbar {\kern \thebaroffset\optbar{\kern -\thebaroffset \Bs}\xspace}

\def\Bcm     {{\ensuremath{\B_\cquark^-}}\xspace}

\def\Bds     {{\ensuremath{\B_{(\squark)}^0}}\xspace}

\def\jpsi     {{\ensuremath{{\PJ\mskip -3mu/\mskip -2mu\Ppsi}}}\xspace}
\def\psitwos  {{\ensuremath{\Ppsi{(2S)}}}\xspace}

\def\Y#1S{\ensuremath{\PUpsilon{(#1S)}}\xspace}

\def\LorLbar     {\kern \thebaroffset\optbar{\kern -\thebaroffset \PLambda}\xspace}

\def\BF         {{\ensuremath{\mathcal{B}}}\xspace}

\newcommand{\decay}[2]{\ensuremath{#1\!\to #2}\xspace} 

\def\to                 {\ensuremath{\rightarrow}\xspace}

\def\AT#1     {\ensuremath{A_{\mathrm{T}}^{#1}}\xspace}           

\def\C#1      {\ensuremath{\mathcal{C}_{#1}}\xspace}                       
\def\Cp#1     {\ensuremath{\mathcal{C}_{#1}^{'}}\xspace}                    
\def\Ceff#1   {\ensuremath{\mathcal{C}_{#1}^{\mathrm{(eff)}}}\xspace}        
\def\Cpeff#1  {\ensuremath{\mathcal{C}_{#1}^{'\mathrm{(eff)}}}\xspace}       
\def\Ope#1    {\ensuremath{\mathcal{O}_{#1}}\xspace}                       
\def\Opep#1   {\ensuremath{\mathcal{O}_{#1}^{'}}\xspace}                    

\newcommand{\nospaceunit}[1]{\ensuremath{\text{#1}}}       
\newcommand{\aunit}[1]{\ensuremath{\text{\,#1}}}       

\newcommand{\tev}{\aunit{Te\kern -0.1em V}\xspace}
\newcommand{\gev}{\aunit{Ge\kern -0.1em V}\xspace}
\newcommand{\mev}{\aunit{Me\kern -0.1em V}\xspace}
\newcommand{\kev}{\aunit{ke\kern -0.1em V}\xspace}
\newcommand{\ev}{\aunit{e\kern -0.1em V}\xspace}
 
\newcommand{\mevc}{\ensuremath{\aunit{Me\kern -0.1em V\!/}c}\xspace}
\newcommand{\gevc}{\ensuremath{\aunit{Ge\kern -0.1em V\!/}c}\xspace}
\newcommand{\mevcc}{\ensuremath{\aunit{Me\kern -0.1em V\!/}c^2}\xspace}
\newcommand{\gevcc}{\ensuremath{\aunit{Ge\kern -0.1em V\!/}c^2}\xspace}

\def\mum  {\ensuremath{\,\upmu\nospaceunit{m}}\xspace}

\def\fb   {\ensuremath{\aunit{fb}}\xspace}
\def\invfb   {\ensuremath{\fb^{-1}}\xspace}

\newcommand{\chisq}{\ensuremath{\chi^2}\xspace}

\def\gsim{{~\raise.15em\hbox{$>$}\kern-.85em
          \lower.35em\hbox{$\sim$}~}\xspace}
\def\lsim{{~\raise.15em\hbox{$<$}\kern-.85em
          \lower.35em\hbox{$\sim$}~}\xspace}

\def\pt         {\ensuremath{p_{\mathrm{T}}}\xspace}

\def\ptot       {\ensuremath{p}\xspace}

\def\evtgen     {\mbox{\textsc{EvtGen}}\xspace}

\def\geant      {\mbox{\textsc{Geant4}}\xspace}

\def\photos     {\mbox{\textsc{Photos}}\xspace}

\def\pythia     {\mbox{\textsc{Pythia}}\xspace}

\def\tell1  {TELL1\xspace}
\def\ukl1   {UKL1\xspace}

\newcommand{\lhcborcid}[1]{\href{https://orcid.org/#1}{\hspace*{0.1em}\raisebox{-0.45ex}{\includegraphics[width=1em]{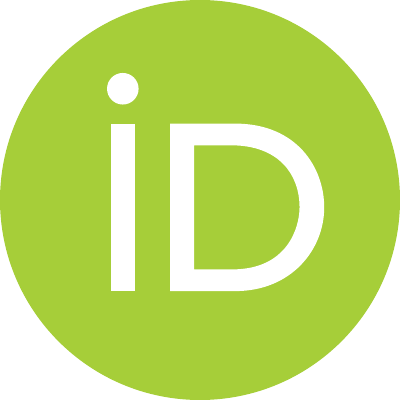}}}}

\usepackage{cite} 
\usepackage{mciteplus}

\def\Bstarz      {{\ensuremath{\B^{*0}}}\xspace}
\def\Bsstar  {{\ensuremath{\B^{*0}_\squark}}\xspace}
\def\BorBs  {{\ensuremath{\B^{0}_{(\squark)}}}\xspace}

\usepackage{longtable} 

\begin{document}

\renewcommand{\thefootnote}{\fnsymbol{footnote}}
\setcounter{footnote}{1}


\begin{titlepage}
\pagenumbering{roman}

\vspace*{-1.5cm}
\centerline{\large EUROPEAN ORGANIZATION FOR NUCLEAR RESEARCH (CERN)}
\vspace*{1.5cm}
\noindent
\begin{tabular*}{\linewidth}{lc@{\extracolsep{\fill}}r@{\extracolsep{0pt}}}
\ifthenelse{\boolean{pdflatex}}
{\vspace*{-1.5cm}\mbox{\!\!\!\includegraphics[width=.14\textwidth]{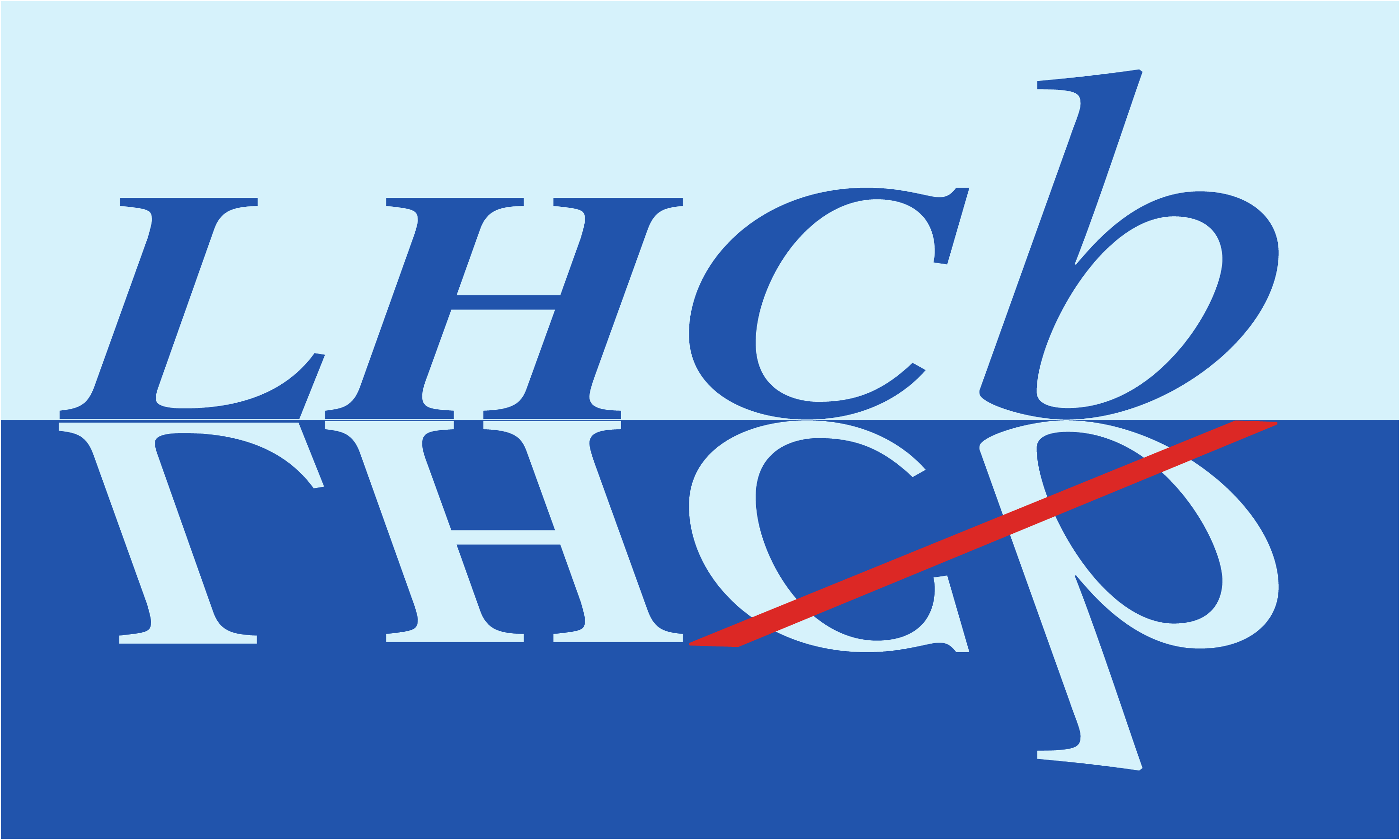}} & &}%
{\vspace*{-1.2cm}\mbox{\!\!\!\includegraphics[width=.12\textwidth]{figs/lhcb-logo.eps}} & &}%
\\
 & & CERN-EP-2023-050 \\  
 & & LHCb-PAPER-2023-004 \\  
 & & Aug 15, 2023 \\
 & & \\
\end{tabular*}

\vspace*{4.0cm}

{\normalfont\bfseries\boldmath\huge
\begin{center}
  \papertitle 
\end{center}
}

\vspace*{2.0cm}

\begin{center}
\paperauthors\footnote{Authors are listed at the end of this paper.}
\end{center}

\vspace{\fill}

\begin{abstract}
  \noindent
  The very rare \decay{\Dstar(2007)^0}{\mumu} decay is searched for by analysing \mbox{\decay{\Bm}{\pim\mumu}} decays. The analysis uses a sample of beauty mesons produced in proton-proton collisions collected with the LHCb detector between 2011 and 2018, corresponding to an integrated luminosity of 9\invfb. The signal signature corresponds to simultaneous peaks in the $\mumu$ and $\pim\mumu$ invariant masses. No evidence for an excess of events over background is observed and an upper limit is set on the branching fraction of the decay at \mbox{${\cal B}(\decay{\Dstar(2007)^0}{\mumu}) < 2.6\times 10^{-8}$} at $90\%$ confidence level. This is the first limit on the branching fraction of \mbox{\decay{\Dstar(2007)^0}{\mumu}} decays and the most stringent limit on $\Dstar(2007)^0$ decays to leptonic final states. The analysis is the first search for a rare charm-meson decay exploiting production via beauty decays.  
\end{abstract}

\vspace*{2.0cm}

\begin{center}
  Published in \href{https://link.springer.com/article/10.1140/epjc/s10052-023-11759-6}{Eur.~Phys.~J.~C 83, 666 (2023)}
\end{center}

\vspace{\fill}

{\footnotesize 
\centerline{\copyright~\papercopyright. \href{\paperlicenceurl}{\paperlicence}.}}
\vspace*{2mm}

\end{titlepage}


\newpage
\setcounter{page}{2}
\mbox{~}


\renewcommand{\thefootnote}{\arabic{footnote}}
\setcounter{footnote}{0}

\cleardoublepage


\pagestyle{plain} 
\setcounter{page}{1}
\pagenumbering{arabic}


\section{Introduction}
\label{sec:Introduction}

Decays of heavy-flavoured hadrons provide a wide range of opportunities to test for possible deviations from Standard Model (SM) expectations. 
The decays of the $\Dz$, $\Bz$ and $\Bs$ mesons into charged lepton pairs are particularly interesting since the chiral structure of the SM weak interaction suppresses their branching fractions, which can be predicted with small theoretical uncertainties~\cite{Bobeth:2013uxa}. 
These features make them highly sensitive to potential contributions from physics beyond the SM.  
Intense activity on the \decay{\BorBs}{\mumu} channels has resulted in the observation of the \decay{\Bs}{\mumu} decay by the LHCb~\cite{LHCb-PAPER-2017-001,LHCb-PAPER-2021-007,LHCb-PAPER-2021-008}, CMS~\cite{Sirunyan:2019xdu} and ATLAS~\cite{Aaboud:2018mst} experiments, and a combined limit on the \decay{\Bd}{\mumu} branching fraction that approaches its SM value~\cite{LHCb-CONF-2020-002}.
The experimental limits on the branching fractions of the decays \decay{\Dz}{\epem}~\cite{Belle:2010ouj}, \decay{\Dz}{\mumu}~\cite{LHCb-PAPER-2013-013,LHCb-PAPER-2022-029}, \decay{\Bds}{\epem}~\cite{LHCb-PAPER-2020-001} and \decay{\Bds}{\tautau}~\cite{LHCb-PAPER-2017-003} are still several orders of magnitude above their SM predictions.

In contrast to the situation for pseudoscalar mesons, the leptonic decays of the excited vector $\Dstarz$, $\Bstarz$ and $\Bsstar$ states have no chiral suppression. 
In an effective field theory approach, these decays involve the same operators as those of the pseudoscalar mesons~\cite{Buchalla:1995vs}.
Therefore, if sufficient experimental precision can be obtained, leptonic decays of the vector states provide a complementary approach to constraining the associated Wilson coefficients~\cite{Wilson:1965zzb}.
The challenge, however, is that the vector mesons can also decay via electromagnetic and (for the $\Dstarz$ meson) strong interactions, which have widths many orders of magnitude larger than those for the weak decays.
Therefore, the branching fractions of these weak decays are strongly suppressed to levels around $10^{-11}$~\cite{Khodjamirian:2015dda,Grinstein:2015aua}, unless there are large enhancements due to physics beyond the SM. 
For the leptonic \Dstarz decays, further suppression by the GIM mechanism~\cite{Glashow:1970gm} reduces the SM prediction for the branching fractions to the ${\cal O}\left(10^{-19}\right)$ level~\cite{Khodjamirian:2015dda,Grinstein:2015aua}.
Leptonic $\Dstarz$, $\Bstarz$ and $\Bsstar$ decays have been considered as a potential probe of physics beyond the SM in Refs.~\cite{Grinstein:2015aua,Khodjamirian:2015dda}, with further investigations of the impact of particular extensions of the SM considered in Refs.~\cite{Xu:2015eev,Sahoo:2016edx,Banerjee:2017upm,Kumar:2017xgl,Kumbhakar:2018uty}.

An upper limit on the branching fraction of \decay{\Dstarz}{\epem} decays has been set by the CMD-3 collaboration, corresponding to $\BF\left(\decay{\Dstarz}{\epem}\right) < 1.7 \times 10^{-6}$ at 90\% confidence level~(CL)~\cite{Shemyakin:2020uye}. 
Throughout this paper, the $\Dstarz$ symbol represents the $\Dstar(2007)^0$ meson, and the inclusion of charge-conjugate processes is implied.
There is no previous published search for \decay{\Dstarz}{\mumu} decays, but in the absence of chiral suppression the branching fractions for \decay{\Dstarz}{\epem} and \decay{\Dstarz}{\mumu} decays are expected to be the same.
As recently discussed in Ref.~\cite{Abudinen:2022dnw}, the copious production of heavy-flavoured hadrons in LHC collisions and the clean experimental signature make it worthwhile to search for \decay{\Dstarz}{\mumu} decays in LHCb data. 
The most promising approach appears to be with the \decay{\Bm}{\Dstarz (\mumu)\, \pim} decay chain since the displaced vertex and exclusive final state provide powerful background rejection capabilities.

This paper describes the first search for the \mbox{\decay{\Dstarz}{\mumu}} decay in the decay chain \mbox{\decay{\Bm}{\Dstarz (\mumu)\,\pim}}. 
The analysis strategy is to reconstruct \decay{\Bm}{\pim\mumu} candidates and apply selection criteria to reduce background. 
The dimuon invariant mass, $m(\mumu)$, and the reconstructed $\B$-candidate invariant mass, $m(\pim\mumu)$, are used as discriminating observables in a fit,
the results of which were not examined until the full analysis procedure had been finalised.
The decay mode \mbox{\decay{\Bm}{\jpsi(\mumu)\, \Km}} is used as a normalisation channel. 
 
The branching fraction for the \decay{\Dstarz}{\mumu} decay is obtained through
\begin{equation}
{\small
\BF\left(\decay{\Dstarz}{\mumu}\right) =
    \frac{N_{\Dstarz\pim}}{N_{\jpsi\Km}} \cdot
    \frac{\varepsilon_{\jpsi\Km}}{\varepsilon_{\Dstarz\pim}} \cdot
    \frac{\BF\left(\decay{\Bm}{\jpsi\Km}\right)}{\BF\left(\decay{\Bm}{\Dstarz\pim}\right)} \cdot
    \BF\left(\decay{\jpsi}{\mumu}\right)\,,
\label{eq:BFnorm}
}
\end{equation}
where $N_{\Dstarz\pim}$ and $\varepsilon_{\Dstarz\pim}$ denote the yield and efficiency for the signal mode, and $N_{\jpsi\Km}$ and $\varepsilon_{\jpsi\Km}$ denote the yield and efficiency for the normalisation mode.
All branching fractions on the right-hand side of Eq.~\eqref{eq:BFnorm} are known~\cite{PDG2022}. The signal- and normalisation-mode yields are determined from invariant-mass fits, and the efficiency ratio is determined from simulation with corrections for data-simulation discrepancies as described below.

\section{Detector and simulation}
\label{sec:Detector} 

The \lhcb detector~\cite{LHCb-DP-2008-001,LHCb-DP-2014-002} is a single-arm forward spectrometer covering the \mbox{pseudorapidity} range $2<\eta <5$, designed for the study of particles containing \bquark or \cquark quarks. 
The detector includes a high-precision tracking system consisting of a silicon-strip vertex detector surrounding the proton-proton ($pp$) interaction region~\cite{LHCb-DP-2014-001}, a large-area silicon-strip detector located upstream of a dipole magnet with a bending power of about $4{\mathrm{\,Tm}}$, and three stations of silicon-strip detectors and straw drift tubes~\cite{LHCb-DP-2013-003,LHCb-DP-2017-001} placed downstream of the magnet.
The tracking system provides a measurement of the momentum, \ptot, of charged particles with
a relative uncertainty that varies from 0.5\% at low momentum to 1.0\% at 200\gevc.
The minimum distance of a track to a primary $pp$ collision vertex~(PV), the impact parameter (IP), 
is measured with a resolution of $(15+29/\pt)\mum$,
where \pt is the component of the momentum transverse to the beam, in\,\gevc.
Different types of charged hadrons are distinguished using information
from two ring-imaging Cherenkov detectors~\cite{LHCb-DP-2012-003}. 
Photons, electrons and hadrons are identified by a calorimeter system consisting of
scintillating-pad and preshower detectors, an electromagnetic
and a hadronic calorimeter. Muons are identified by a
system composed of alternating layers of iron and multiwire
proportional chambers~\cite{LHCb-DP-2012-002}.
The online event selection is performed by a trigger~\cite{LHCb-DP-2012-004,LHCB-DP-2019-001}, 
which consists of a hardware stage, based on information from the calorimeter and muon
systems, followed by a two-level software stage, which reconstructs the full event.

Simulation is used to tune the event selection procedure, to model the shape of the dimuon and \B-candidate invariant-mass distributions and to estimate efficiencies accounting for the effects of the detector acceptance, reconstruction and selection criteria. 
In the simulation, $pp$ collisions are generated using \pythia~\cite{Sjostrand:2007gs} with a specific \lhcb configuration~\cite{LHCb-PROC-2010-056}. 
Decays of unstable particles are described by \evtgen~\cite{Lange:2001uf}, in which final-state radiation is generated using \photos~\cite{davidson2015photos}.
The interaction of the generated particles with the detector, and its response, are implemented using the \geant toolkit~\cite{Allison:2006ve, *Agostinelli:2002hh} as described in Ref.~\cite{LHCb-PROC-2011-006}. 
The underlying $pp$ interaction is reused multiple times, with an independently generated signal decay for each~\cite{LHCb-DP-2018-004}.

The \B candidates reconstructed in simulation are weighted to correct for discrepancies between data and simulation associated with particle-identification efficiency, track-reconstruction efficiency and hardware trigger efficiency. Additional corrections are applied to account for discrepancies in \B-production kinematics and event track multiplicity.
After these weights are applied,
the simulated distributions of all variables used in the analysis are in good agreement with the data.

The particle identification efficiencies are determined from data using unbiased samples of identified charged particles from \decay{\jpsi}{\mumu} and \decay{\Dstarp}{\Dz(\Km\pip)\,\pip} decays~\cite{LHCb-DP-2018-001}. 
The efficiencies are determined in intervals of track momentum, track pseudorapidity and event track multiplicity to match the properties of the signal and normalisation modes. 
Differences between data and simulation associated with the track reconstruction efficiency are corrected using samples of \decay{\jpsi}{\mumu} decays~\cite{LHCb-DP-2013-002}.
The hardware trigger efficiency is determined, using independently selected \mbox{\decay{\Bm}{\jpsi(\mumu)\,\Km}} decays, 
in intervals of the transverse momenta of the $\mup$ and the $\mun$~tracks~\cite{LHCb-PUB-2014-039}.
Discrepancies associated with the \B-production kinematics and the event track multiplicity are corrected using a multivariate algorithm~\cite{Rogozhnikov:2016bdp}, which is trained using \mbox{\decay{\Bm}{\jpsi(\mumu)\,\Km}} decays in background-subtracted data and simulation. 
As a cross-check, an alternative algorithm is trained using the number of tracks involved in the PV reconstruction as additional input, and consistent results are obtained.

\section{Candidate selection} 
\label{sec:Selection} 

Events are required to pass a hardware trigger that selects events containing at least one muon with high transverse momentum. 
The $\pt$ threshold is between $1.2$ and $2.2\gevc$ depending on the data-taking period. 
In the subsequent software trigger, at least one muon is required to have $\pt > 0.8\gevc$ and impact parameter greater than $100\mum$ with respect to all PVs in the event. 
The tracks of two or more of the final-state particles are required to form a vertex that is significantly displaced from any PV. 
A multivariate algorithm~\cite{BBDT,LHCb-PROC-2015-018} 
is used to identify secondary vertices consistent with the decay of a \bquark hadron.

The \B~candidates are formed from pairs of well-reconstructed oppositely charged tracks that are identified as muons, combined with an additional track that is identified as either a charged pion or a charged kaon for \decay{\Bm}{\Dstarz\pim} or \decay{\Bm}{\jpsi\Km} candidates, respectively. 
Each track is required to have a good fit quality, a low probability of overlapping with any other track in the event, $\pt > 300\mevc$ and to be inconsistent with originating directly from any PV. 
To ensure reliable particle identification, each track is further required to have $5 < p < 100\gevc$ and $ 1.9 < \eta < 4.9$, and events are required to contain fewer than 300 tracks.
The \B~candidates are required to have a good-quality vertex fit.
The \B~candidates are each associated with the PV giving the smallest value of $\chi^2_{\rm IP}$, which is defined as the difference in the vertex-fit \chisq of the PV reconstructed with and without the candidate.
Each \B~candidate must be consistent with originating from its associated PV, with a momentum vector aligned with the direction between the primary and \B-decay vertices.

Each \B candidate is required to have an invariant mass in the range \mbox{$5180 < m(h^-\mumu) < 6000 \mevcc$}, where $h^-$ represents the pion or kaon in the signal or normalisation mode, respectively. 
The dimuon invariant mass is calculated from the outcome of a kinematic fit in which the \B-candidate invariant mass is constrained to the known \Bm~mass~\cite{PDG2022} and the momentum vector is constrained to be consistent with the line of flight between the PV and the decay vertex, thereby improving the resolution. 
The dimuon invariant mass is required to be in the range $1900 < m(\mumu) < 2100 \mevcc$ for the signal mode and $3000 < m(\mumu) < 3200 \mevcc$ for the normalisation mode. 
The expected signal resolution corresponds to about $19\mevcc$ in the \B-candidate invariant mass and about $7\mevcc$ in the dimuon invariant mass.

Combinatorial background arising from random combinations of tracks is suppressed using a multivariate classifier. 
A boosted decision tree~(BDT) algorithm~\cite{Breiman,AdaBoost}, as implemented in the TMVA toolkit~\cite{Hocker:2007ht,*TMVA4}, is trained using supervised learning with ten-fold cross validation~\cite{stone1974cross} to achieve an unbiased classifier response.  
The BDT classifier is trained to identify the \decay{\Bm}{\pim\mumu} signal candidates irrespective of dimuon invariant mass.
The signal sample used for the BDT training comprises simulated nonresonant \decay{\Bm}{\pim\mumu} decays.  
The background training sample comprises data from the upper sideband in the regions $5500 < m(\pim\mumu) < 7000\mevcc$ with vetoes to remove candidates containing \decay{\jpsi}{\mumu}, \decay{\psitwos}{\mumu} and \decay{\Bcm}{\pim\mumu} decays. 
As no particle identification information is used in the classifier, it can be applied to both the \decay{\Bm}{\Dstarz\pim} and the \decay{\Bm}{\jpsi\Km} samples.
 The features of the data used to classify \B~candidates are: the \pt and IP of the pion and muon tracks; the flight distance, \pt and vertex quality of the \B~candidate; a measure of the isolation of the \B~candidate~\cite{LHCb-PAPER-2015-025}; the number of tracks used to reconstruct the PV that is most consistent with being the origin of the \B~candidate; and the absolute difference between the momenta of the muons. 

The requirements on the BDT output and the particle identification of the charged pion are optimised simultaneously to obtain the best signal sensitivity. 
The optimisation is based on pseudoexperiments. 
Ensembles of pseudoexperiments are generated in a grid of points corresponding to different selection requirements. 
The generation includes only the expected background components: nonresonant \decay{\Bm}{\pim \mumu} decays; \decay{\Bm}{\Km \mumu} decays, where the \Km is mistakenly identified as a \pim; and combinatorial background. 
No \decay{\Bm}{\Dstarz(\mumu)\,\pim} signal is included as this component is not expected to be significant. 
The distributions of the nonresonant backgrounds are modelled using simulation, whilst those of the combinatorial background are modelled using sideband data. 
The sideband excludes \B~candidates inside a signal region of $\pm 3$ times the expected dimuon invariant-mass signal resolution, centred at the known $\Dstarz$ mass~\cite{PDG2022}. 
The expected yields for all background components are determined at each grid point by fitting a background-only model to the sideband and interpolating the results.
Each pseudoexperiment is fitted in the same way as data (see Sec.~\ref{sec:fit}).

The minimised figure of merit is defined as the expected uncertainty on the signal yield divided by the signal efficiency, as this is proportional to the expected upper limit in the absence of signal. 
The expected uncertainty on the signal yield is the width of the distribution of residuals in the ensemble.
Alternative figures of merit are found to give similar working points.
With the optimal requirements, the classifier has a combinatorial background rejection power of $99.3\%$, whilst retaining $84.1\%$ of \decay{\Bm}{\Dstarz(\mumu)\,\pim} decays. 
The particle identification requirements have a signal efficiency of about $86\%$, with a charged kaon misidentification rate of about $3\%$.
Each selected event contains only a single \B~candidate. 
As the classifier separates \B~decay signal from combinatorial background, high-purity samples of \decay{\Bm}{\jpsi(\mumu)\,\Km} decays are obtained using the same classifier requirements, when requiring a positively identified kaon.

Backgrounds from partially reconstructed decays such as \mbox{\decay{\Bzb}{\rhoz(\pip\pim)\,\mumu}} and \decay{\Bsb}{f_0(\pip\pim)\,\mumu} for the signal mode and \decay{\Bzb}{\jpsi(\mumu)\,\Kstarzb(\Km\pip)} for the normalisation mode have a reconstructed \B-candidate invariant mass that lies more than $100\mevcc$ below the known \Bm~mass~\cite{PDG2022}.
Consequently, these background components lie outside of the fit range used in the analysis.  
No significant contributions from other partially reconstructed backgrounds are expected within the dimuon invariant-mass fit ranges for the signal and normalisation modes. 
This is true for decays with a missing photon, such as \mbox{$\decay{\Bm} {\eta^{(\prime)}(\mumu\gamma)\,\pim}$}, or with one or more missing neutrinos,
such as \mbox{\decay{\Bm}{\Dz(h^{-}\mup\neum)\,\mun\neumb}}.
Contributions from hadronic backgrounds such as \mbox{\decay{\Bm}{\pim\pip\pim}} decays, where two pions are mistakenly identified as muons are also estimated using simulation~\cite{LHCb-PAPER-2019-017} and are found to be negligible. 

Unavoidably, a considerable number of nonresonant \decay{\Bm}{\pim\mumu} decays~\cite{LHCb-PAPER-2015-035} survive the signal selection. 
This background has a \B-candidate invariant-mass signature identical to the signal, but has a smoothly varying dimuon invariant-mass distribution over the fit range~\cite{Bouchard:2013eph}.  
Possible interference effects between the signal and nonresonant \decay{\Bm}{\pim \mumu} decays are neglected due to the narrow $\Dstarz$ meson width.
Due to the imperfect pion-kaon separation, misidentified \decay{\Bm}{\Km \mumu} decays also form an important background. 
This background favours lower \B-candidate invariant-mass values due to the kaon-pion mass difference, and has a smoothly varying dimuon invariant-mass distribution over the fit range.

For the normalisation mode, the background from misidentified \decay{\Bm}{\jpsi(\mumu)\,\pim} decays is small due to its comparatively low branching fraction and the additional suppression from the kaon identification requirement. Nonetheless, due to the large sample size, it is considered in the fit for the normalisation channel.

\section{Invariant-mass fits}
\label{sec:fit}

The \decay{\Bm}{\Dstarz(\mumu)\,\pim} yield is determined from a two-dimensional extended unbinned maximum-likelihood fit to the $m(\mumu)$ and $m(\pim\mumu)$ distributions of the selected \B~candidates. 
The fit model includes four components: signal \mbox{\decay{\Bm}{\Dstarz\pim}} decays, nonresonant \decay{\Bm}{\pim \mumu} decays, misidentified nonresonant \decay{\Bm}{\Km \mumu} decays and combinatorial background. 
The dimuon and the \B-candidate invariant-mass distributions for each component are modelled using empirical analytical functions. 
The models for the signal and the two nonresonant backgrounds are validated using simulation. 
No significant dependencies between the two fit variables are observed in simulation or sideband data, so they are treated as uncorrelated.  

The signal dimuon invariant-mass distribution is modelled using a Gaussian function with power-law tails on both sides of the distribution~\cite{Skwarnicki:1986xj}. 
The signal \B-candidate invariant-mass distribution is modelled using the sum of a Gaussian function and a Gaussian function with power-law tails.
The tail parameters are fixed to the values obtained from simulation. 
The signal dimuon and \B-candidate invariant-mass models each include a global shift of peak position and a global scaling factor for the width of the distribution, relative to the values found in simulation.

For the two nonresonant backgrounds, the dimuon invariant-mass distributions are described by first-order polynomial functions, with slope parameters fixed to values obtained from simulation. 
The \B-candidate invariant-mass distribution for the nonresonant $\decay{\Bm}{\pim \mumu}$ background is modelled with the same function used for the signal, whilst the \mbox{$\decay{\Bm}{\Km \mumu}$} background is modelled with a Gaussian function with power-law tails. 
The tail parameters are fixed to the values obtained from simulation.
The \B-candidate invariant-mass models for the two nonresonant background components share the global peak position shift and width scaling factor with the signal \B-candidate invariant-mass model.

For the combinatorial background, the dimuon and the \B-candidate invariant-mass distributions are modelled using a first-order polynomial function and an exponential function, respectively. 
The dimuon and the \B-candidate invariant-mass slopes are allowed to vary in the fit to data. 

\begin{figure}[!tb]
\centering
\includegraphics[width=0.50\textwidth]{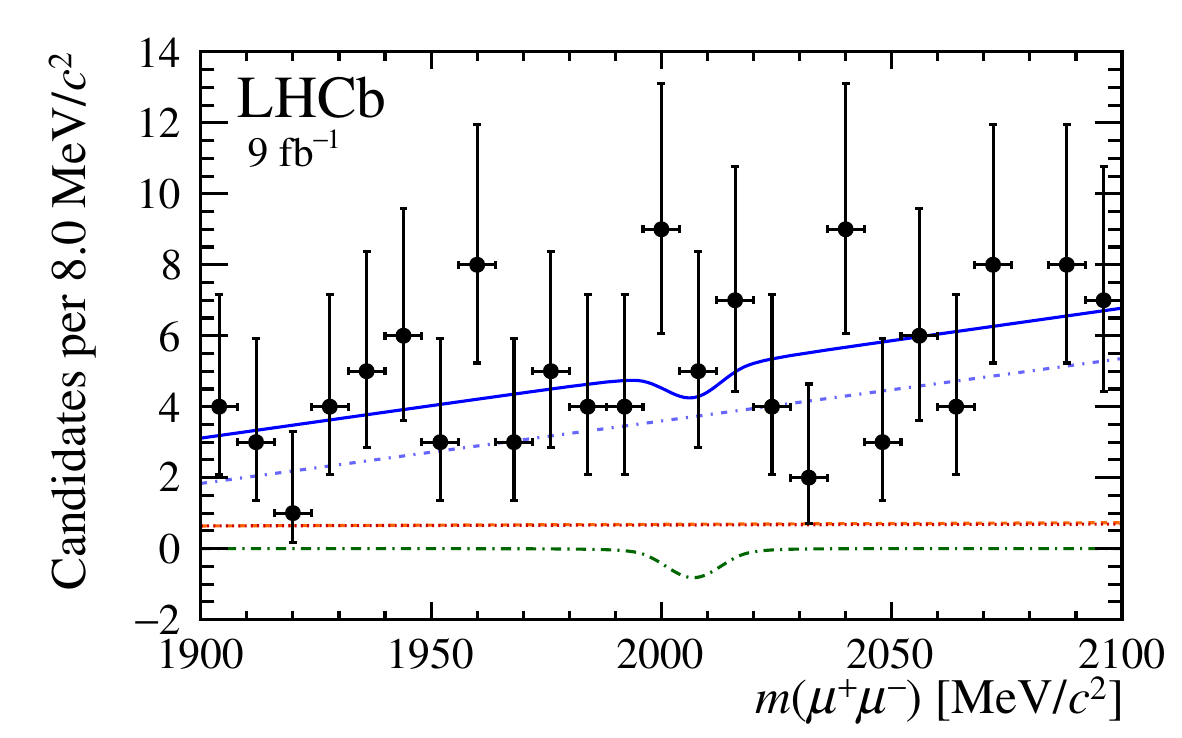}\hfill
\includegraphics[width=0.50\textwidth]{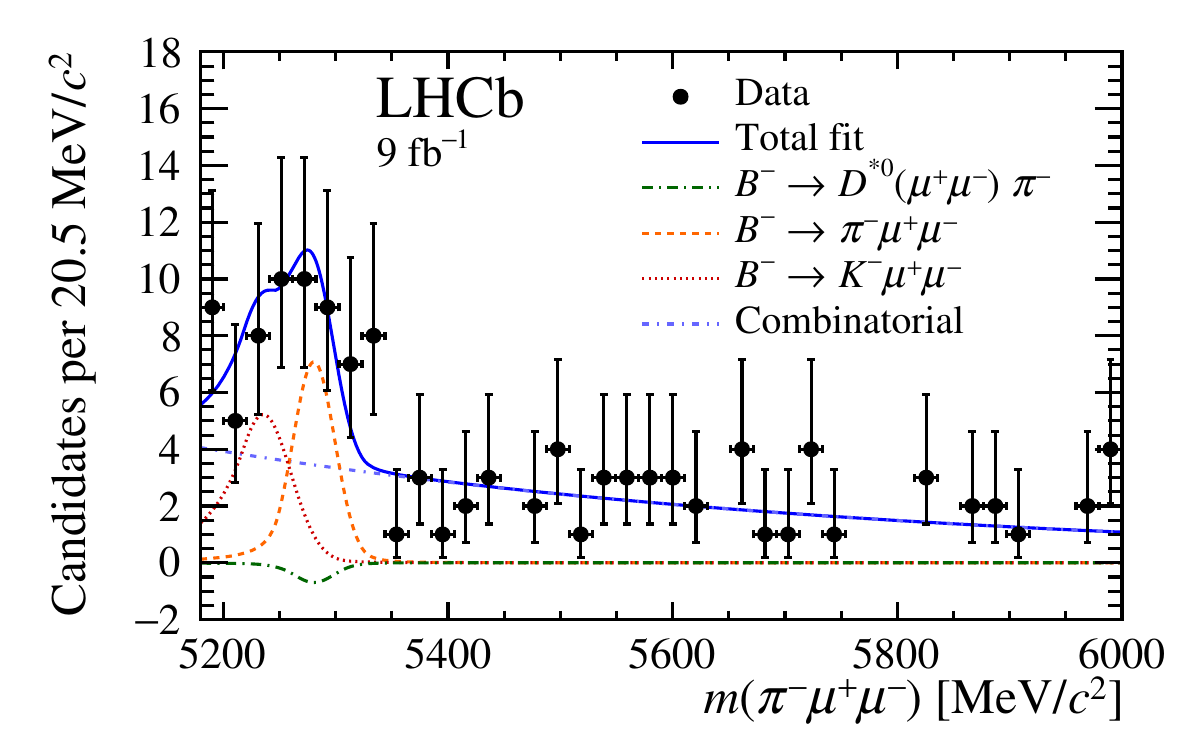}
\caption{
    Reconstructed (left) $\mumu$ and (right) $\pim\mumu$ invariant-mass distributions for the selected \mbox{\decay{\Bm}{\Dstarz(\mumu)\,\pim}} candidates, with results of the fit superimposed. 
}
\label{fig:mass:dstar} 
\end{figure}

\begin{figure}[tb!]
    \centering
    \includegraphics[width=0.6\textwidth]{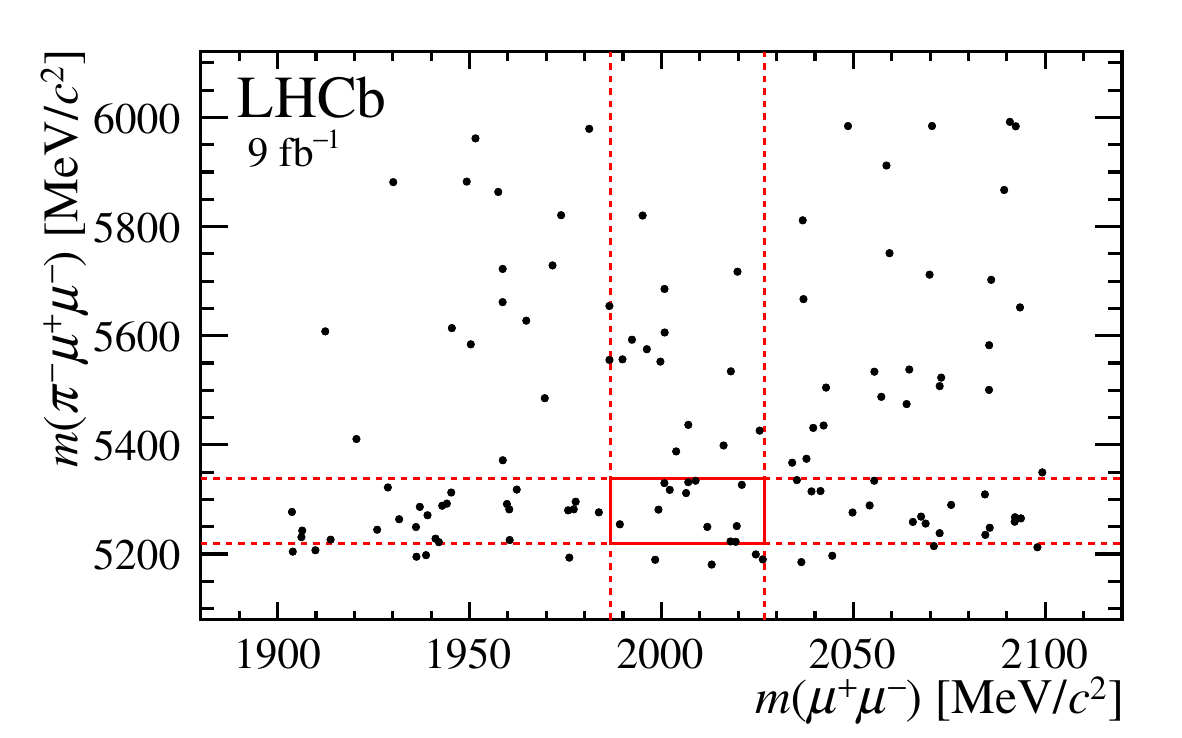}
    \caption{
    Two-dimensional distribution of $\mumu$ invariant mass versus $\pim\mumu$ invariant mass for the selected \decay{\Bm}{\Dstarz(\mumu)\,\pim} candidates. 
    The red box corresponds to a range of about $\pm3\sigma$ around the expected signal peak position in each dimension. 
    }
    \label{fig:unblinded-data-2Dscatter}
\end{figure}

In total, the fit includes six free parameters: the yields for each component and the two parameters of the combinatorial background model. 
The global peak position shift and width scaling factor for each of the dimuon and \B-candidate invariant-mass models are constrained in the fit to be consistent with values obtained from fits to the \decay{\Bm}{\jpsi(\mumu)\,\Km} candidates described below. 
Figure~\ref{fig:mass:dstar} shows the dimuon and \B-candidate invariant-mass distributions of selected \decay{\Bm}{\Dstarz\pim} candidates, with results of the fit superimposed.
Figure~\ref{fig:unblinded-data-2Dscatter} shows the two-dimensional distribution of selected candidates.
The fit favours a slightly negative \decay{\Bm}{\Dstarz\pim} yield, which is attributed to a downward fluctuation of the background in the region close to the \Bm and \Dstarz meson masses. Table~\ref{tab:fitresults_fixed} summarises the yields obtained from the fit.

{\renewcommand{\arraystretch}{1.2}
\begin{table}[tb]
    \centering
    \caption{Yields obtained from the fit to data described in the text, with statistical uncertainties only.}    
    \begin{tabular}{cr@{$\,\pm\,$}l}
       \hline
       Component   & \multicolumn{2}{c}{Yield}\\
       \hline
       \decay{\Bm}{\Dstarz(\mumu)\,\pim}   & $-2$  & $3$ \\
       \decay{\Bm}{\pim\mup\mun}                                 & $ 17  $  & $ 7  $ \\
       \decay{\Bm}{\Km\mup\mun}                                  & $ 17  $  & $ 8  $ \\
       Combinatorial\ bkg.\                                            & $ 90  $  & $13  $ \\\hline
    \end{tabular}
    \label{tab:fitresults_fixed}
\end{table}
}

The \decay{\Bm}{\jpsi(\mumu)\,\Km} yield is determined from a one-dimensional extended unbinned maximum-likelihood fit to the $m(\Km\mumu)$ distribution. 
The normalisation mode also serves to obtain the constraints on the correction factors that account for discrepancies between data and simulation in the signal peak position and width. 
Therefore, an additional maximum-likelihood fit is performed to the $m(\mumu)$ distribution to determine these factors. 
The $B$-candidate invariant-mass and dimuon invariant-mass fits to the normalisation mode are independent of each other. 
A two-dimensional fit is not used since possible correlations in the tail regions of the two observables could cause fit bias due to the large sample size. 

The fits to the \decay{\Bm}{\jpsi(\mumu)\,\Km} candidates include three components: \mbox{\decay{\Bm}{\jpsi\Km}} decays, misidentified \decay{\Bm}{\jpsi\pim} decays, and combinatorial background. 
The dimuon and the $B$-candidate invariant-mass distributions for each component are modelled using empirical functions. 
For the \mbox{\decay{\Bm}{\jpsi\Km}} component, the dimuon and \B-candidate invariant-mass distributions are modelled with the same functions used for the signal \mbox{\decay{\Bm}{\Dstarz\pim}} component in the signal-mode fit. 
For the misidentified \mbox{\decay{\Bm}{\jpsi\pim}} component, both the \B-candidate and the dimuon invariant-mass distributions are modelled using a Gaussian function with power-law tails on both sides. 
The parameters for the \decay{\Bm}{\jpsi\Km} and the misidentified \decay{\Bm}{\jpsi\pim} components are determined from simulation, but a global peak position shift and a width scaling factor are allowed to vary in the fits to data.

Combinatorial background is modelled using an exponential function in the \B-candidate invariant-mass distribution and a first-order polynomial function in the dimuon invariant-mass distribution. 
The \B-candidate and the dimuon invariant-mass slopes are allowed to vary in the fits. 

\begin{figure}[!tb]
\centering

\includegraphics[width=0.49\linewidth]{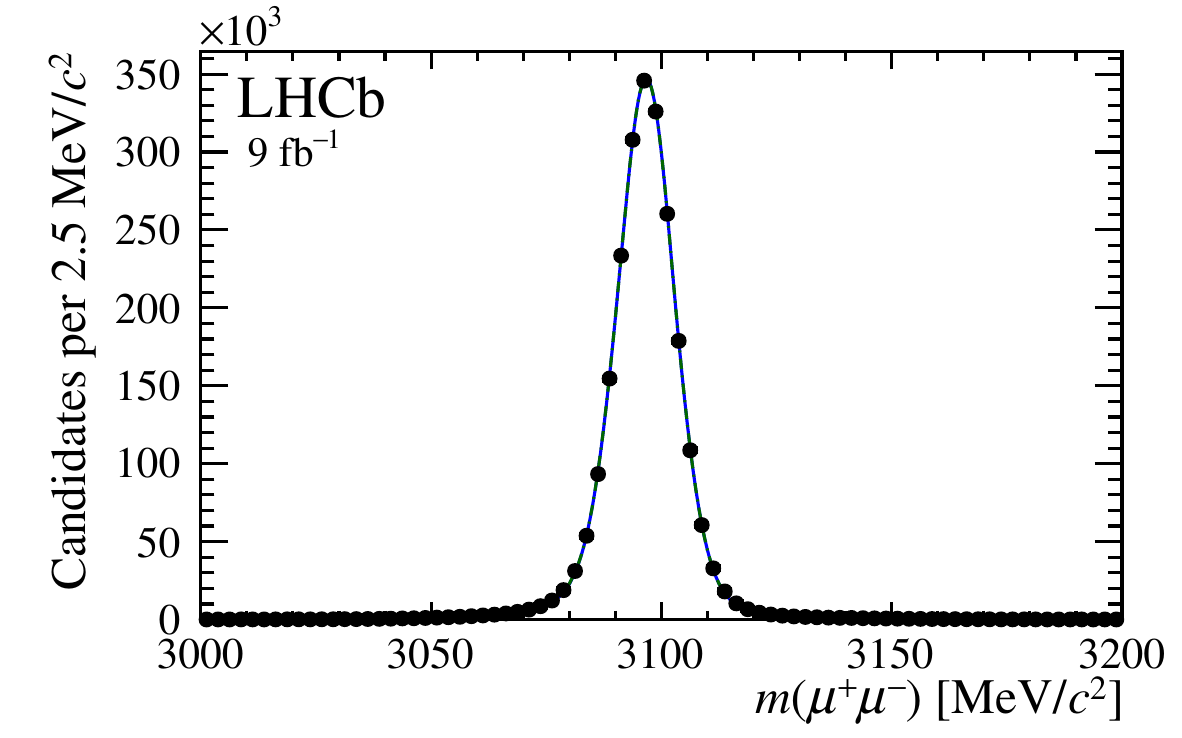}
\hfill
\includegraphics[width=0.49\linewidth]{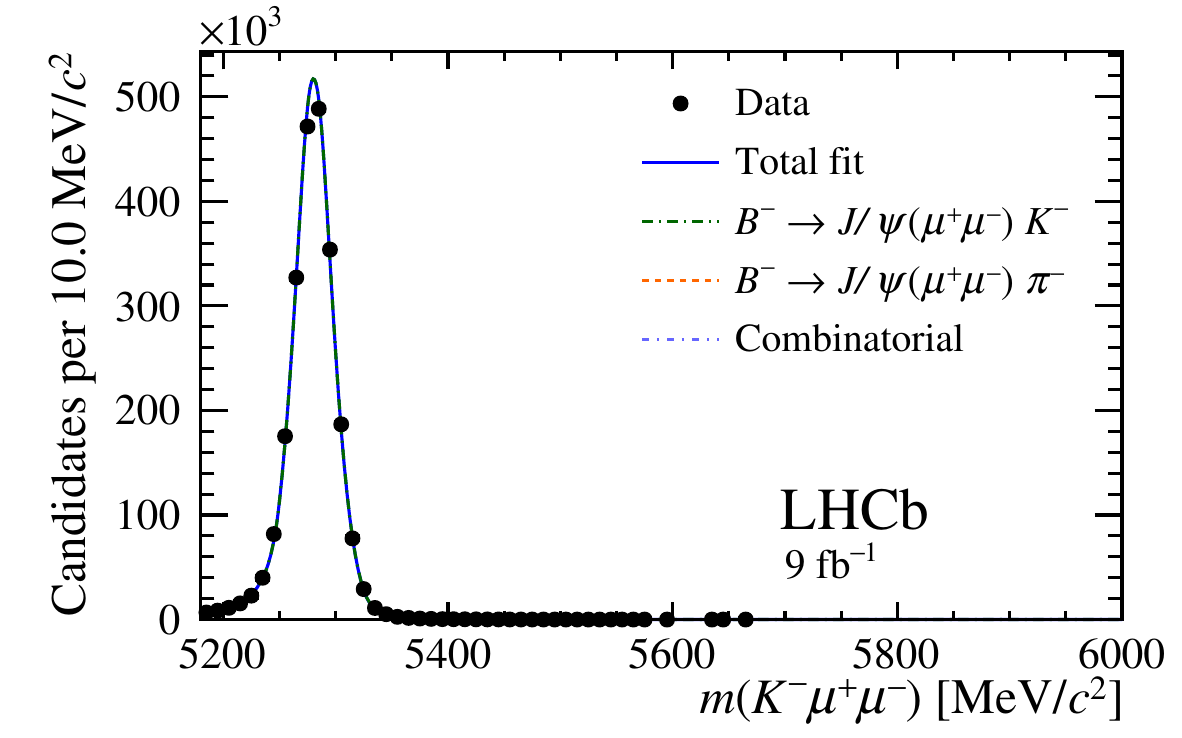} 
\caption{
    Reconstructed (left) $\mumu$ and (right) $\Km\mumu$ invariant-mass distributions for the selected \mbox{\decay{\Bm}{\jpsi(\mumu)\,\Km}} candidates, with results of the fits superimposed.
    The contributions from misidentified \decay{\Bm}{\jpsi(\mumu)\pim} decays and combinatorial background are  negligible and thus barely visible.
}
\label{fig:mass:jpsi} 
\end{figure}

In total, each one-dimensional fit model includes six free parameters: the yields for each component, the global peak shift and width scaling factor and the model parameter of the combinatorial background. 
Figure~\ref{fig:mass:jpsi} shows the dimuon and \B-candidate invariant-mass distributions of selected \mbox{\decay{\Bm}{\jpsi\Km}} candidates. 
The dimuon and the $B$-candidate invariant-mass fits converge to \decay{\Bm}{\jpsi\Km} yield values that are consistent within $0.4\%$; the $B$-candidate invariant-mass fit converges to $(2316 \pm2)\times 10^{3}$ decays, where the uncertainty is statistical only. 
The contributions from misidentified \decay{\Bm}{\jpsi\pim} decays and combinatorial background are found to be negligible. 

\section{Systematic uncertainties and results} 
\label{sec:Results} 

To obtain the branching fraction of the \decay{\Dstarz}{\mumu} decay, the signal yield in the fit described in Sec.~\ref{sec:fit} is parameterised in terms of $\BF\left(\decay{\Dstarz}{\mumu}\right)$ using Eq.~\eqref{eq:BFnorm}. 
The values of the branching fractions on the right-hand side of the equation, the efficiency ratio and the normalisation yield are allowed to vary within Gaussian constraints to account for the uncertainties on these inputs. 
Table~\ref{tab:syst_constrs} summarises the constraints. 
The widths of the constraints correspond to the statistical and systematic uncertainties added in quadrature.
For the measured branching fractions, the values from Ref.~\cite{PDG2022} are used.
The efficiency ratio ${\varepsilon_{\jpsi\Km}}/{\varepsilon_{\Dstarz\pim}}$ is obtained from simulation, accounting for the geometrical acceptance of the detector as well as effects related to the triggering, reconstruction and selection of the \B~candidates.
The normalisation yield is obtained from the fits described in the previous section.

{\renewcommand{\arraystretch}{1.2}
\begin{table}[tb!]
    \centering
    \caption{Input parameters used in the estimation of the \decay{\Dstarz}{\mumu} branching fraction. The uncertainties correspond to the statistical and systematic uncertainties added in quadrature.}  
    \begin{tabular}{c r@{$\,\pm\,$}l} 
    \hline
    Parameter & \multicolumn{2}{c}{Value} \\\hline
        $\BF\left(\decay{\Bm}{\jpsi\Km}\right)$ &  $(10.20$ & $0.19)\times 10^{-4}$~\cite{PDG2022} \\
       $\BF\left(\decay{\Bm}{\Dstarz\pim}\right)$  & $(4.90$ & $0.17)\times 10^{-3}$~\cite{PDG2022} \\
       $\BF\left(\decay{\jpsi}{\mumu}\right)$ & $(59.61$ & $0.33)\times 10^{-3}$~\cite{PDG2022} \\
       ${\varepsilon_{\jpsi\Km}}/{\varepsilon_{\Dstarz\pim}}$ & $1.21 $ & $0.03$ \\
       $N_{\jpsi\Km}$ & $(2316$ & $8)\times 10^{3}$\\
       \hline
    \end{tabular}
    \label{tab:syst_constrs}
\end{table}
}

The uncertainty on the efficiency ratio takes into account the simulation sample size, uncertainties on the weights applied to the simulation 
and the matching between reconstructed and generated particles in the simulation. 
The systematic uncertainties associated with the weights are evaluated by varying all weights within their uncertainties and by varying the binning used to estimate them. 
The systematic uncertainty associated with the multivariate weighting algorithm is evaluated by comparing the results obtained with the default and alternative algorithms~(see Sec.~\ref{sec:Detector}). 
The systematic uncertainty associated with the matching between reconstructed and generated particles in the simulation is evaluated by comparing the efficiencies obtained including or excluding \B~candidates for which one or more decay products are not correctly matched. 
All variations are made consistently for the signal and normalisation modes to avoid an overestimation of the uncertainty on the efficiency ratio. 
The systematic effects associated with the uncertainties on the weights cancel out almost fully in the determination of the efficiency ratio. 
The effect of the matching between reconstructed and generated particles dominates the uncertainty on the efficiency ratio, but  has no significant impact on the modelled invariant-mass distributions.

The systematic uncertainty on the yield of the normalisation mode is evaluated by comparing the yields obtained in four different approaches: 
the baseline fit to the \B-candidate invariant-mass distribution; 
a fit replacing the \mbox{\decay{\Bm}{\jpsi\, \Km}} shape by a Hypatia function~\cite{MartinezSantos:2013ltf}; 
the baseline fit to the dimuon invariant-mass distribution; 
and a fit to the dimuon invariant-mass distribution replacing the \mbox{\decay{\Bm}{\jpsi\, \Km}} shape by the sum of a Gaussian function and a Gaussian function with power-law tails on both sides of the distribution. 
The largest difference is assigned as the systematic uncertainty.
The difference between the fit models is also used to assign a systematic uncertainty on the global peak position shifts and width scaling factors for the signal mode. 

\begin{figure}[tb!]
    \centering
    \includegraphics[width=0.6\textwidth]{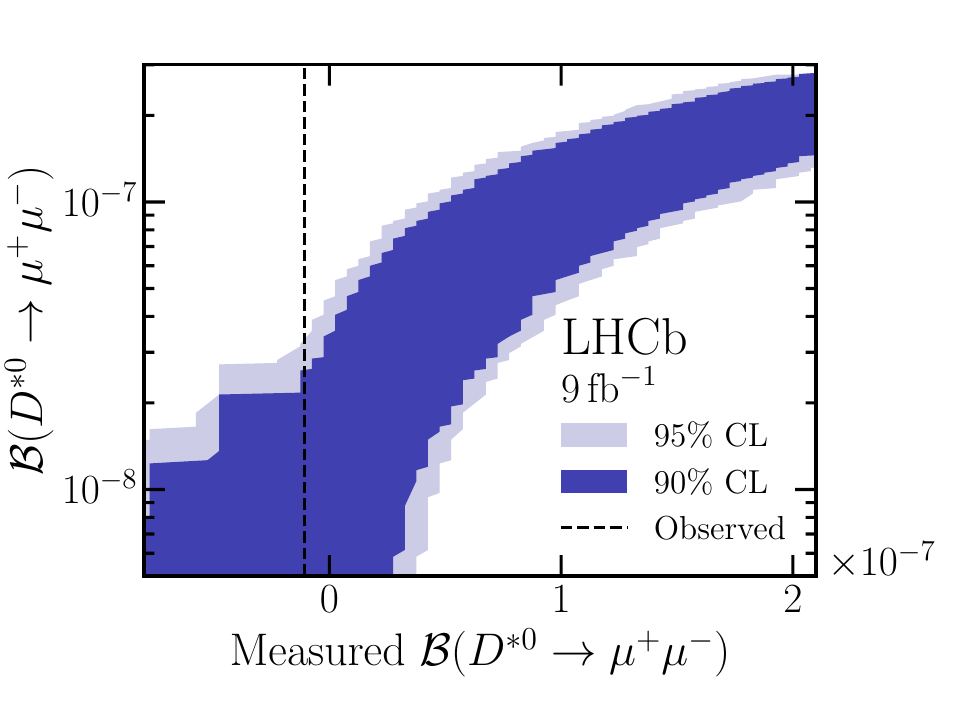}
    \caption{
    Confidence belts generated using pseudoexperiments according to the Feldman--Cousins prescription~\cite{Feldman:1997qc}.
    The vertical black line shows the result of the fit to data.
    }
    \label{fig:data-unblinded-conf-band-Btrue}
\end{figure}

Including all constraints, the fit to data yields 
\begin{equation*}
\BF\left(\decay{\Dstarz}{\mumu}\right) = (-1.06\pm 1.85)\times 10^{-8}\,,    
\end{equation*}
where statistical and systematic uncertainties are combined.
An upper limit on the branching fraction is obtained following the Feldmann--Cousins prescription~\cite{Feldman:1997qc}: pseudoexperiments are generated for various values of the branching fraction and the resulting distribution of measured branching fractions is used to form confidence belts. Figure~\ref{fig:data-unblinded-conf-band-Btrue} shows confidence belts at $90\%$ and $95\%$ CL. 
The result obtained from the fit yields 
  \begin{equation*}
    \BF(\decay{\Dstarz}{\mumu}) < 2.6\, (3.4) \times 10^{-8} \text{ at 90 (95)\% CL}\,.
  \end{equation*}
The procedure is repeated fixing the nuisance parameters to their central values to assess the impact of the systematic uncertainties.
In this case, with statistical uncertainty only, the fitted branching fraction is $(-1.10\pm1.72)\times 10^{-8}$ and the obtained 90 (95)\%~CL upper limit is $2.3\, (3.2) \times 10^{-8}$, indicating that the result is statistically limited.
As further checks the procedure is repeated restricting the signal yield to positive values, or using alternative signal dimuon and \B-candidate invariant-mass shapes.
No significant change in the upper limit is found in any of these checks.

\section{Summary} 
\label{sec:Summary} 

A search for the \decay{\Dstarz}{\mumu} decay is performed by analysing \decay{\Bm}{\pim\mumu} decays.  
The analysis uses a data sample corresponding to an integrated luminosity of $9\invfb$ collected with the LHCb experiment in $pp$ collisions between 2011 and 2018. This is the first search for a rare charm-meson decay exploiting its production in beauty-meson decays. No excess with respect to the
background-only hypothesis is observed and an upper limit of $\BF(\decay{\Dstarz}{\mumu}) < 2.6 \times 10^{-8}$ at $90\%$~CL is set. 
This measurement is the first limit on the branching fraction of \decay{\Dstarz}{\mumu} decays and the most stringent limit on \Dstarz decays to leptonic final states.

\section*{Acknowledgements}

\noindent We express our gratitude to our colleagues in the CERN
accelerator departments for the excellent performance of the LHC. We
thank the technical and administrative staff at the LHCb
institutes.
We acknowledge support from CERN and from the national agencies:
CAPES, CNPq, FAPERJ and FINEP (Brazil); 
MOST and NSFC (China); 
CNRS/IN2P3 (France); 
BMBF, DFG and MPG (Germany); 
INFN (Italy); 
NWO (Netherlands); 
MNiSW and NCN (Poland); 
MEN/IFA (Romania); 
MICINN (Spain); 
SNSF and SER (Switzerland); 
NASU (Ukraine); 
STFC (United Kingdom); 
DOE NP and NSF (USA).
We acknowledge the computing resources that are provided by CERN, IN2P3
(France), KIT and DESY (Germany), INFN (Italy), SURF (Netherlands),
PIC (Spain), GridPP (United Kingdom), 
CSCS (Switzerland), IFIN-HH (Romania), CBPF (Brazil),
Polish WLCG  (Poland) and NERSC (USA).
We are indebted to the communities behind the multiple open-source
software packages on which we depend.
Individual groups or members have received support from
ARC and ARDC (Australia);
Minciencias (Colombia);
AvH Foundation (Germany);
EPLANET, Marie Sk\l{}odowska-Curie Actions and ERC (European Union);
A*MIDEX, ANR, IPhU and Labex P2IO, and R\'{e}gion Auvergne-Rh\^{o}ne-Alpes (France);
Key Research Program of Frontier Sciences of CAS, CAS PIFI, CAS CCEPP, 
Fundamental Research Funds for the Central Universities, 
and Sci. \& Tech. Program of Guangzhou (China);
GVA, XuntaGal, GENCAT, Inditex and InTalent (Spain);
SRC (Sweden);
the Leverhulme Trust, the Royal Society
 and UKRI (United Kingdom).



\addcontentsline{toc}{section}{References}
\bibliographystyle{LHCb}
\bibliography{main,standard,LHCb-PAPER,LHCb-CONF,LHCb-DP,LHCb-TDR}

\newpage
\centerline
{\large\bf LHCb collaboration}
\begin
{flushleft}
\small
R.~Aaij$^{32}$\lhcborcid{0000-0003-0533-1952},
A.S.W.~Abdelmotteleb$^{51}$\lhcborcid{0000-0001-7905-0542},
C.~Abellan~Beteta$^{45}$,
F.~Abudin{\'e}n$^{51}$\lhcborcid{0000-0002-6737-3528},
T.~Ackernley$^{55}$\lhcborcid{0000-0002-5951-3498},
B.~Adeva$^{41}$\lhcborcid{0000-0001-9756-3712},
M.~Adinolfi$^{49}$\lhcborcid{0000-0002-1326-1264},
P.~Adlarson$^{77}$\lhcborcid{0000-0001-6280-3851},
H.~Afsharnia$^{9}$,
C.~Agapopoulou$^{43}$\lhcborcid{0000-0002-2368-0147},
C.A.~Aidala$^{78}$\lhcborcid{0000-0001-9540-4988}
% ...

{\footnotesize \it

% ...
$^{9}$Universit{\'e} Clermont Auvergne, CNRS/IN2P3, LPC, Clermont-Ferrand, France\\
% ...
$^{32}$Nikhef National Institute for Subatomic Physics, Amsterdam, Netherlands\\
% ...
$^{41}$Instituto Galego de F{\'\i}sica de Altas Enerx{\'\i}as (IGFAE), Universidade de Santiago de Compostela, Santiago de Compostela, Spain\\
% ...
$^{43}$European Organization for Nuclear Research (CERN), Geneva, Switzerland\\
% ...
$^{45}$Physik-Institut, Universit{\"a}t Z{\"u}rich, Z{\"u}rich, Switzerland\\
% ...
$^{49}$H.H. Wills Physics Laboratory, University of Bristol, Bristol, United Kingdom\\
% ...
$^{51}$Department of Physics, University of Warwick, Coventry, United Kingdom\\
% ...
$^{54}$School of Physics and Astronomy, University of Glasgow, Glasgow, United Kingdom\\
$^{55}$Oliver Lodge Laboratory, University of Liverpool, Liverpool, United Kingdom\\
% ...
$^{63}$Syracuse University, Syracuse, NY, United States\\
% ...
$^{77}$Department of Physics and Astronomy, Uppsala University, Uppsala, Sweden, associated to $^{54}$\\
$^{78}$University of Michigan, Ann Arbor, MI, United States, associated to $^{63}$\\
\bigskip
% ...
}
\end{flushleft}

\end{document}